\documentclass[twocolumn,superscriptaddress,showpacs,amsfonts]{revtex4-1}
\pdfoutput=1
\usepackage{graphicx}
\usepackage{bm}
\usepackage{amsmath}
\usepackage{amsfonts}
\usepackage{amssymb}
\usepackage{epsfig}
\usepackage{color}
\usepackage{dsfont}

\newcommand{\abs}[1]{\left\vert#1\right\vert}

\newcommand{\bra}[1]{\left\langle#1\right\vert}
\newcommand{\ket}[1]{\vert#1\rangle}

\newcommand\braket[2]{\langle#1|#2\rangle}

\begin{document}
%Guiding
\title{Effects of geometry on spin-orbit Kramers states in semiconducting nanorings}

\author{G. Francica}

\affiliation{CNR-SPIN, c/o Universit\`a di
Salerno, I-84084 Fisciano (Salerno), Italy}

\author{P. Gentile}

\affiliation{CNR-SPIN, c/o Universit\`a di
Salerno, I-84084 Fisciano (Salerno), Italy}

\author{M. Cuoco}

\affiliation{CNR-SPIN, c/o Universit\`a di
Salerno, I-84084 Fisciano (Salerno), Italy}

%\pacs{03.65.Vf}{Phases: geometric; dynamic or topological}
%\pacs{05.30.Rt}{Quantum phase transition}
%\pacs{03.65.-w}{Quantum mechanics}

\begin{abstract}
{The holonomic manipulation of spin-orbital degenerate states, encoded in the Kramers doublet of narrow semiconducting channels with spin-orbit interaction, is shown to be intimately intertwined with the geometrical shape of the nanostructures. The presence of doubly degenerate states is not sufficient to guarantee a non-trivial mixing by only changing the Rashba spin-orbit coupling. We demonstrate that in nanoscale quantum rings the combination of arbitrary inhomogeneous curvature and adiabatic variation of the spin-orbit amplitude, {\it e.g.} through electric-field gating, can be generally employed to get non-trivial combinations of the degenerate states. Shape symmetries of the nanostructure act to constrain the adiabatic quantum evolution.
While for circular rings the geometric phase is not generated along a non-cyclic path in the parameters space, remarkably, for generic mirror-symmetric shape deformed rings the spin-orbit driving can lead to a series of dynamical quantum phase transitions. We explicitly show this occurrence and propose a route to detect such topological transitions by measuring a variation of the electron tunneling amplitude into the semiconducting channel.}
\end{abstract}

\maketitle

{\it Introduction --} The interest in geometric phases in quantum mechanics has been turned on by the seminal works of Berry and Simon~\cite{berry84,simon83}.
For an adiabatic and cyclic evolution, it is known that the quantum state acquires a phase factor that depends only on the path in the parameters manifold. Indeed, the geometric phase is due to a holonomy for the given fibre bundle~\cite{simon83}. For degenerate quantum systems the phase factor is non-Abelian, as firstly pointed out by Wilczek and Zee for a cyclic evolution~\cite{wilczek84}. Geometric phases have been then generalized for non-cyclic paths~\cite{bhandari88,mukunda93,mostafazadeh99,kult06} by showing their intimate relation with the geometric structures of the projective Hilbert space~\cite{page87,anandan90,book}.

One of the major challenges in the context of quantum processing points to the achievement of novel mechanisms and devices which are based on non-Abelian geometric phases, as they are robust to local perturbations and can provide means for performing universal quantum information processing ~\cite{zanardi99,sjoqvist15} with noteworthy perspectives towards the realization of geometric quantum computation~\cite{unanyan99,duan01,fuentes-guridi02,recati02,solinas03} .
In this framework, low-dimensional semiconductors with inversion asymmetry can be particularly attractive~\cite{nowack07,nadjperge10,frolov13,petta13,manchon15} because the spin-orbit (SO) interaction~\cite{spinorbit1,spinorbit2,spinorbit3} offers a tantalizing prospect of an all-electrical control over the electron spin in absence of a magnetic field. Furthermore, the electrical manipulation preserves time-reversal symmetry which is crucial to guarantee degenerate Kramers pair configurations by which a qubit can be in principle encoded and quantum processed.

Along this direction, schemes to perform non-Abelian holonomic operations on the electron spin have been mainly focusing on time-dependent electrostatic confining potentials realized through moving quantum dots~\cite{golovach10}, also including closed loop trajectories~\cite{cadez14,kregar16}. Although experimental realizations of moving quantum dots have been successfully achieved~\cite{sanada13},
the efficiency of such quantum engineering is sensitively dependent on the concomitant dynamical control of the confining electrostatic potential and the strength of the inversion symmetry breaking via the SO coupling. It would be then highly beneficial to develop alternative paths which can provide expanded functionalities and increase the phase space for quantum manipulation.

In this Letter, we aim at demonstrating how spin-orbital quantum states, encoded in the Kramers doublet, can be engineered in nanoscopic semiconducting channels using the combined effects of geometric curvature and Rashba spin-orbit coupling. The potential of this union has already led to augmented paths for the design of topological states \cite{gentile15,saarikoski2015,ying16,reynoso17,scopigno18}
and electron-transport \cite{frustaglia04,bercioux05,nagasawa2013}. Such effects mainly arise due to the fundamental role of nanoscale shaping in narrow SO coupled semiconducting channels acting as a source of spatial dependent spin-torque controlling both the electron spin-orientation and its spin-phase through non-trivial spin windings~\cite{ying16,saarikoski2015}.
Here, we show that asymmetrically shaped nanostructures can generally lead to non-trivial mixing of the states forming the Kramers doublet by an adiabatic driving of the spin-orbit amplitude, {\it e.g.} through electric-field gating.
%On the other hand, the spin manipulation should be affected by shape
%deformations. Indeed the nonuniform curvature of SO coupled %semiconducting gives rise to spin textures with topological features %in space~\cite{ying16}.

Moreover, since the geometric curvature of SO coupled nanostructures can generate spin-texture with nontrivial topological windings in real space, it is relevant to evaluate whether such geometrical resource can lead to quantum driven topological phase transitions. Recently, a subtle interconnection between topological phase changeover and dynamical quantum phase transitions (DQPT)~\cite{heyl18} has been put into limelight in a variety of quantum systems~\cite{heyl18,hickey14}, including low-dimensional topological systems~\cite{vajna15}. Remarkably, for a given symmetry class, a topological invariant, represented by a momentum space winding number of the Pancharatnam geometric phase, can be introduced and employed for characterizing DQPTs~\cite{budich16}.
%In this work, we aim at demonstrating the fundamental role played by %the geometrical shape of semiconducting narrow channels in the %presence of Rashba SO coupling for achieving non-Abelian geometric %phase manipulation.
In this paper, for suitable shaped nanostructures with mirror symmetry, we find that the amplitude's variation of the Rashba SO coupling can lead to a series of dynamical topological phase transitions. A controlled amplification of the Rashba interaction is feasable by electric-field gating~\cite{nitta97,liang12} as it is a consolitated practice in a variety of semiconducting nanostructures including small quantum rings. Advancements in the design of small nanoring with few electrons \cite{lorke00,keyser03,fuhrer01} and different shapes make our proposal accessible in laboratory. To this end, we discuss possible setups for detecting the spin-orbit driven quantum topological transitions.
%In this paper, we demonstrate that nonuniform curvature allows to achieve a geometric phase in the non-cyclic adiabatic evolution whose character is intimately tied to the spatial symmetries of the nanostructure.
%While for a circle a geometric phase cannot be produced, a nonuniform curvature allows to get a nonzero geometric phase.
%Indeed, a quantized phase is originated if the spatial profile of the quantum ring owes %mirror symmetry. We find that $\pi$-jumps in the geometric phase are observed every time %the system undergoes through a DQPT with a changing in the topological features of the %quantum state. We demonstrate that an arbitrary geometrical profile of the nanostructure %uniquely leads to non-Abelian phase connection in the Kramers degenerate space.

%
\begin{figure}
\includegraphics[width=0.5\textwidth]{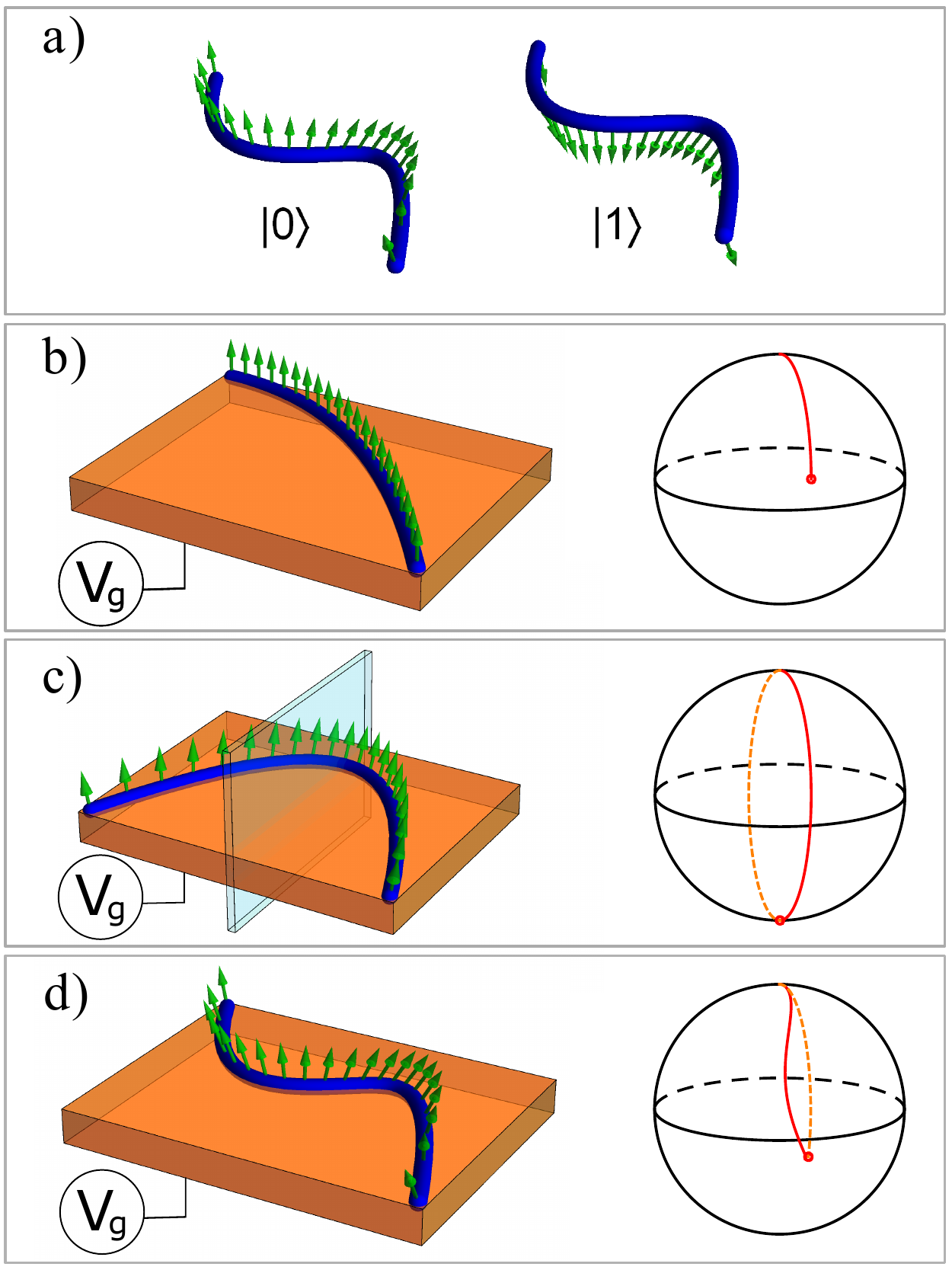}
\caption{(a) schematic representation of the spin-orbital 
Kramers pairs with time-reversal symmetric spin-texture along a curved spin-orbit coupled nanochannel (blue line). 
Sketches of solid-state platform with applied gate voltage $V_g$ for tuning the strength of the Rashba spin-orbit coupling. A schematic view of the adiabatic evolution of spin-orbital $\ket{0}$ and $\ket{1}$ states, associated to the spin-textures in (a), is represented on an effective Bloch sphere with the corresponding trajectory 
(red solid line) for the case of (b)~constant curvature, (c)~mirror symmetric and generically \textrm{d)}~shaped nanostructure. The dotted line (orange) indicates the geodesic for closing the path for the case of a non-cyclic evolution and provide the geometric phase.}
\label{fig:DQPT}
\end{figure}

{\it Model system --} We consider a system of electrons propagating in a narrow semiconducting channel lying in the $xy$-plane and forming a spatial profile with a nontrivial spatial curvature. The shape of the narrow nanostructure can be generally specified by introducing two unit vectors $\hat{\mathcal T}(s)$ and $\hat{\mathcal N}(s)$, which are tangent and normal to the spatial profile at a given position labelled by the curvilinear coordinate $s$.
The spatial dependent spin-orbit coupling is expressed via two local Pauli matrices, comoving with the electrons as they propagate along $s$, expressed by
$\sigma_{\mathcal N}(s) = {\boldsymbol \sigma} \cdot \hat{\mathcal N}(s)$ and $\sigma_{\mathcal T}(s) = {\boldsymbol \sigma} \cdot \hat{\mathcal T}(s)$,
where the $\boldsymbol{\sigma}$'s are the usual Pauli matrices.
$\hat{\mathcal{T}}$ and $\hat{\mathcal{N}}$ are related via the Frenet-Serret type equation $\partial_s \hat {\mathcal{N}} = \kappa\,\hat {\mathcal{T}}$ with $\kappa$ being the local curvature.
%Using the Frenet-Serret  equation $\partial_s \hat{\mathcal T}(s) \equiv \kappa(s) %\hat{\mathcal N}(s)$, it then follows that the angle $\varphi(s)= - \int^s %\kappa(s^{\prime}) d s^{\prime}$ is entirely determined by the local curvature.
%The normal and the tangential directions to the wire can be expressed in terms of a polar angle $f$  as  $\hat  N = (\cos f,\sin f,0 )$ and $\hat T = (\sin f,-\cos f,0 )$. By considering the arclength $s$, the two versors $\hat N$ and $\hat T$ satisfy the Frenet-Serret (FS) type equation $\partial_s \hat N = k \hat T$ where $k = - \partial_s f$ is the local curvature.
A Rashba SO coupling arises due to the inversion symmetry breaking and can be tuned by an applied electric field transverse to the plane of the nanostructure. An effective model description that is able to capture the combination of Rashba spin-orbit coupling and geometrical curvature is given by the
%an electron in the conduction band can be described by the %$\mathbf k \cdot \mathbf p$
Hamiltonian~\cite{gentile13,ortix15,gentile15,zhang07}
%The normal and the tangential directions to the wire at the position specified by the arclength $s$ can be expressed in terms of the polar angle $f(s)$  as  $\hat  N(s) = (\cos f(s),\sin f(s),0 )$ and $\hat T(s) = (\sin f(s),-\cos f(s),0 )$ respectively. The two versors satisfy the Frenet-Serret type equation $\partial_s \hat N = K \hat T$ where $K = - \partial_s f$ is the local curvature.
%The system is described by the Hamiltonian~\cite{PRB16}
%\begin{equation}\label{Hami}
%\hat H= \frac{\hat p^2}{2m} -\frac{\hbar \alpha}{2m} (\hat\sigma_N(\hat s) \hat p +  \hat p\hat\sigma_N(\hat s))
%\end{equation}
\begin{equation}\label{Hami}
H= -\frac{ \hbar^2}{2m}\partial_s^2 +i\frac{\hbar^2 \alpha}{2m} (\sigma_N(s) \partial_s +  \partial_s \sigma_N(s))
\end{equation}
where $m$ is the effective mass of the charge carrier, $\alpha$ is the Rashba SO coupling strength and $\sigma_N $ is the normal component of the spin with respect to the nanostructure geometrical profile.
Since we are interested in assessing the role of the geometry to set the character of the Kramers pairs quantum evolution, we consider different types of spatial profiles focusing on the ensuing symmetries.
%For this aim,
%the nanostructure can be characterized by its spatial symmetries.
The circular quantum nanoring is a highly symmetric case and, thus, it represents an ideal reference with uniform curvature and invariance under continuous rotation around the axis perpendicular to the electron orbital plane. A deviation from the circular shape can bring to two possible paths of shape deformations: a first one preserving few specific point group symmetries of the quantum loop and a second direction corresponding to nanoring with an arbitrary shape.
%with respect to the angle, and shape deformations will lead to lose some symmetries. For %our purposes
We consider the class $C_1$ of nanoring geometry with point symmetry transformations including the rotation around the $z$ axis (or in-plane inversion), $P:(x,y,0)\mapsto (-x,-y,0)$, and a subclass $C_2\subset C_1$ for which there is also a mirror symmetry with respect to the reflection $P_y:(x,y,0)\mapsto (x,-y,0)$ or equivalently $P_x$. The computational analysis is performed by numerically solving the model Hamiltonian (1) and deals with the case of nanoring with circular, elliptical or generic asymmetric shape. The elliptical configuration is a representative and paradigmatic example of planar inversion and mirror symmetric profiles exhibiting an inhomogeneous curvature.
% belongs to the class $C_2$, and deformations allow to break reflection symmetry and to go in $C_1\setminus C_2$.
In general, for any geometrical shape, the model Hamiltonian is symmetric with respect to the time-reversal transformation $\Theta$, so that the energies $E$ are two-fold degenerate and the eigenstates arise in Kramers pairs.
%The time-reversal transformation can be represented by the antiunitary operator $\Theta = i \sigma_y K$ where $K$ is the operation of complex conjugation.
For inversion symmetric profiles of the nanoring, the Hamiltonian is also invariant under the transformation $M = \sigma_z P $, so that a Kramers pair $(\ket{E_+},\ket{E_-})$ can be classified such that $\Theta \ket{E_+} = \ket{E_-}$ and $M \ket{E_\pm}  = \pm \ket{E_\pm}$, and the Hilbert space is the direct sum $\mathcal{H} = \mathcal{H}_+ \oplus \mathcal{H}_-$ with the invariant subspace $\mathcal{H}_\pm$ being spanned by $\{\ket{E_{\pm}}\}_E$.
%$H$ is invariant under the time reversal transformation $\hat\Theta=i\sigma_y K$ and the energies $E$ are two-fold degenerate. The system is symmetric with respect to the transformation $M= \sigma_z P_z = \sigma_z e^{i p L/2\hbar}$. The two eigenstates $\ket{E_{1}}$ and $\ket{E_{2}}$ with energy $E$ are chosen so that $\hat \Theta \ket{E_1} = \ket{E_2}$ and $M \ket{E_{1,2}}=\pm \ket{E_{1,2}}$, the Hilbert space is the direct sum of the two invariant subspace $\mathcal{H} = \mathcal{H}_+ \oplus \mathcal{H}_-$. The subspace $\mathcal{H}_\pm$ is spanned by $\{\ket{E_{1,2}}\}_E$.

It has been shown in Refs.~\cite{ying16,saarikoski2015} that the electron spin orientation manifests topological features expressed by windings of the spin direction along the spatial profile.
%In particular, for a quantum state with spin component $\boldsymbol\psi$, the spin orientation with respect to the Frenet-Serret frame is given by $\langle \boldsymbol{\sigma} \rangle =  \boldsymbol\psi^\dag \boldsymbol \sigma\boldsymbol\psi$.
In particular, for a nonuniform curvature the spin orientation $\langle \boldsymbol\sigma \rangle$ with respect to the Frenet-Serret frame displays spin textures which correspond to loops on the Frenet-Serret-Bloch sphere, and are characterized by windings around the normal and the out-of-plane directions.
With the same spirit, another topological feature of the state can be introduced by considering the relative phase $\chi=\arg \boldsymbol\psi^\dag(L)\boldsymbol\psi(0)=2\pi w$ where $\boldsymbol\psi$ indicated the spin configuration of the state, $w$ is an integer and $L$ is the length of the nanoring. %and $\boldsymbol\psi$ is the spin component of the state.
%When the system is driven across these regions by changing the SO Rashba coupling $\alpha$  a non trivial non-equilibrium dynamics will emerge.
These topological aspects are relevant and play an important role in setting the geometric phase produced in the quantum evolution.

{\it Spin-orbit driven quantum geometric phase --} Let us start by providing few general considerations about the phase acquired by a given state for non-cyclic adiabatic time evolution which is achieved by changing sufficiently slowly the Rashba SO coupling from the initial value $\alpha_{in}$ into the final one $\alpha$ in a given nanoring. %On a general ground, the time-evolution in an energy eigenspace is determined by the Wilczek-Zee connection~\cite{wilczek84}, at least of a dynamical phase.
If the system is initially prepared in an eigenstate $\ket{E_a(\alpha_{in})}$, the adiabatically evolved state $\ket{\Psi_a(\alpha)}$ will be a linear combination of the degenerate eigenstates, {\it i.e.} $\ket{\Psi_a(\alpha)} = \sum U_{b,a}(\alpha) \ket{E_b(\alpha)}$ at least of an irrelevant dynamical phase factor, so that the evolved state acquires a non-Abelian phase factor $U(\alpha)$ which depends on the basis chosen $\{\ket{E_a(\alpha)}\}_a$ with the eigenstates smooth in $\alpha$.
The holonomic contribution can be taken in account for an open path in the projective Hilbert space by a unitary matrix $U_g(\alpha)$ having gauge-invariant eigenvalues~\cite{mostafazadeh99,kult06}, which are  $e^{\pm i\gamma}$ due to the time reversal symmetry. %for the system under consideration.

By considering the initial state $\ket{E_+(\alpha_{in})}$, the evolution occurs in the invariant subspace $\mathcal{H}_+$ and the evolved state can be expressed as $\ket{\Psi_+(\alpha)}=e^{i\gamma}\ket{\tilde E_+(\alpha)}$, where the state $\ket{\tilde E_+(\alpha)}$ is defined from $\ket{E_+(\alpha)}$ by fixing the gauge so that $\arg\braket{E_+(\alpha_{in})}{\tilde E_+(\alpha)}=0$. The phase $\gamma$ can be so viewed as a Pancharatnam phase~\cite{bhandari88,mukunda93}, and can be expressed as the integral of the Berry-Simon connection evaluated along the curved electronic nanochannel

\begin{equation}
\nonumber
\gamma = \iint \boldsymbol\psi^\dag i \partial_\alpha \boldsymbol\psi %\text{d} s \text{d}\alpha
\end{equation}
\noindent or equivalently in a more compact form $\gamma = \int {\mathcal{A}}=\int \langle {\tilde E}_+ {\text{d}} {\tilde E}_+ \rangle$, where $\boldsymbol\psi$ is the spin component of the state $\ket{\tilde E_+}$, which satisfies the differential equation~\cite{ying16} $i\partial_s \boldsymbol \psi(s) = \tilde{Q}(s) \boldsymbol\psi(s) $, with $\tilde{Q}(s) = -\alpha \sigma_N(s) - c_0$, and $c_0^2=\frac{2 m}{\hbar^2} E+\alpha^2$.

During the evolution, due to the continuous change in the spin direction at the point $s$, one can identify a local geometric phase $\gamma_B(s)$~\cite{bhandari88,mukunda93}. This part is uniquely related to the changing in the local spin direction during the evolution%({\it i.e.} invariant under local gauge transformations $\boldsymbol\psi(s) \mapsto e^{i\phi(s)}\boldsymbol\psi(s)$)
, and it is half the solid angle swept by the geodesic closure of the path spanned by the local spin direction during the evolution.

Then, the phase $\gamma$ can be expressed as

\begin{equation}
\gamma =  \int_0^L \gamma_B \rho \text{d} s +\Delta \lambda
\end{equation}

\noindent where $\Delta \lambda=-\int_0^L \varphi(s)  \rho ds$, $\varphi=\arg(\boldsymbol \psi^\dag(s,\alpha_{in}) \boldsymbol \psi(s,\alpha))$ being the total phase, and $\rho=\boldsymbol \psi^\dag\boldsymbol \psi$ is the spatially constant electron density.

Starting from the expression of the phase $\gamma$ we observe that $\Delta \lambda$ is related to an eventual changing in the winding number $\Delta w = w(\alpha)-w(\alpha_{in})$ through the general equation
\begin{equation}
\Delta \lambda = \pi\Delta w - \frac{1}{2}\int_0^{L}(\varphi(s) + \varphi(-s))\rho \text{d} s \,.
\end{equation}

Let us consider the geometric phase for various spatial symmetries of the nanoring.

The circular quantum ring is a special case of mirror symmetric profile because, due to the constant radius, it has also invariance under rotation around the axis perpendicular to the ring plane. Such symmetry aspect is further constraining the possible values of the acquired phase along the adiabatic evolution.
Indeed, under a rotation of an angle $\theta$ with respect to the $z$-axis we have $\boldsymbol\psi\mapsto e^{i\left (w-\frac{1}{2} - \frac{\sigma_z}{2}\right)\theta}\boldsymbol\psi$ where $w$ is the winding number previously introduced, from which the orthogonality condition cannot be reached, {\it i.e.} $\tilde G(\alpha)=\braket{E_+(\alpha_{in})}{\tilde E_+(\alpha)}= \boldsymbol\psi^\dag(s,\alpha_{in}) \boldsymbol\psi(s,\alpha) L >0$.
Furthermore, exploiting the same argumentation it is straightforward to show that there is only another state, indicated with $\ket{ E'_+(\alpha_{in})}$, having a non-zero overlap with respect to the state $\ket{\tilde E_+(\alpha)}$, so that

\begin{equation}
\ket{\tilde E_+(\alpha)} = \tilde G(\alpha) \ket{E_+(\alpha_{in})} + \sqrt{1-\tilde G^2(\alpha)}  \ket{ E'_+(\alpha_{in})}
\end{equation}

This parameterizes an arc of great circle in the space of the unit vectors $\mathcal S(\mathcal H)$, so that the adiabatic evolution moves along a geodesic in the projective Hilbert space, for which the Pancharatnam phase is zero, {\it i.e.} $\gamma=0$~\cite{anandan90,book}. In particular $\mathcal L_{FS}=\arccos \tilde G(\alpha)$ is its length calculated with respect to the Fubini-Study metric~\cite{page87}.

{\it Topological phase transition and non-Abelian phase --} Proceeding with our analysis for mirror symmetric nanostructures, we have that $\tilde{Q}(s)=\tilde{Q}^\dag(-s)$ so that $\boldsymbol\psi(-s) = K \boldsymbol\psi(s)$, with $K$ being the complex conjugate operator. Taking into account such symmetry relations, one can show that the sum of all the local geometric phases is zero $\int \gamma_B\, \rho \text{d} s=0$, and the local total phase is odd $\varphi(s)=-\varphi(-s)$, so that a variation in the phase $\gamma$ can be uniquely related to a change in the winding $w$, {\it i.e.} $\gamma = \Delta \chi /2 =\pi\Delta w$.
The value of $\Delta w$ is such that the condition $\arg\braket{E_+(\alpha_{in})}{\tilde E_+(\alpha)}=0$ is satisfied.
We have

\begin{equation}
\braket{E_+(\alpha_{in})}{\tilde E_+(\alpha)}=\int \boldsymbol \psi^\dag (s,\alpha_{in}) \boldsymbol\psi(s,\alpha)  \text{d} s
\end{equation}

Hence, we observe that $\boldsymbol\psi^\dag (s,\alpha_{in}) \boldsymbol\psi(s,\alpha) = r e^{i\Delta\chi/2}e^{i\phi}$ with $r\geq0$, so that under a reflection with respect to the $xz$ plane, $r e^{i\Delta\chi/2}e^{i\phi} \mapsto r e^{i\Delta\chi/2}e^{-i\phi}$. We then obtain

\begin{equation}
\braket{E_+(\alpha_{in})}{\tilde E_+(\alpha)}=2e^{i\Delta\chi/2}\int_0^{\frac{L}{2}} r \cos(\phi) \text{d} s
\end{equation}

\noindent and consequently $e^{i\Delta\chi/2}=e^{i\pi\Delta w}=\pm1$ in order to keep $\arg\braket{E_+(\alpha_{in})}{\tilde E_+(\alpha)}=0$ during the evolution. %We note that the change in sign occurs when the evolved state becomes orthogonal with respect to the initial state.

%The geometric local contribution to the phase, which is only related on the changing in the local spin direction during the evolution (i.e. invariant under local gauge transformations $\boldsymbol\psi(s) \mapsto e^{i\phi(s)}\boldsymbol\psi(s)$), is given by the local geometric phase~\cite{bhandari88,mukunda93} indicated with $\gamma_B(s)$. It is half the solid angle subtended by the curve spanned by the local spin direction closed with the geodesic arc.

%where we have defined the local total phase $\varphi(s)=\arg(\boldsymbol \psi^\dag(s,\alpha_{in}) \boldsymbol \psi(s,\alpha))$.
%with $\boldsymbol \psi$ being the spin component of the state $\ket{\tilde E_+}$.

%{\it Topological phase transition and non-Abelian phase --} Let us apply the previous observations to the various geometrical profiles of the nanoring by explicitly evaluating the geometric phase. For the ideal case of a circular ring, since the eigenstates have a uniform spin profile, the resulting adiabatic evolution gives a geodesic arc on the projective Hilbert space so that a geometric phase is not generated.
~%\cite{appendix}. %$\mathcal{P}(\mathcal{H})$.
%On the other hand, a nonuniform curvature can move the evolution away from the geodesic.
%For shapes with reflection symmetry (i.e. class $C_2$),  the sum of all the local geometric phases is zero $\int \gamma_B\, \rho ds=0$, and the total phase is odd $\varphi(s)=-\varphi(-s)$, so that a variation in the phase $\gamma$ can be uniquely related to a change in the winding $w$, i.e. $\gamma = \pi\Delta w$.

Then for shapes with reflection symmetry, the winding number parity $\Delta w(\mod 2)$ can change at given values  $\alpha_n$ of the Rashba SO coupling for which the final evolved state becomes orthogonal with respect to the initial one.
%~\cite{appendix}.
The eigenvalue $e^{i\gamma}$ gives a sign factor, whose value can be obtained from the following connection %evaluated on the curve $\ket{\tilde E_+}$ reads
\begin{equation}
\mathcal{A} =  \sum_n \pi \delta(\alpha-\alpha_n) \text{d}\alpha
\end{equation}
% from which we see that the geometric-phase is Abelian $U_g=\pm1$.
By explicitly evaluating the holonomy for an elliptical nanoring, we find that the adiabatic evolution keeps the evolved state in phase with the initial one until the orthogonality is reached and the phase $\gamma$ undergoes a $\pi$-jump (see Fig.~\ref{fig:DQPT}).

\begin{figure}[!t]
\includegraphics[width=0.98\columnwidth]{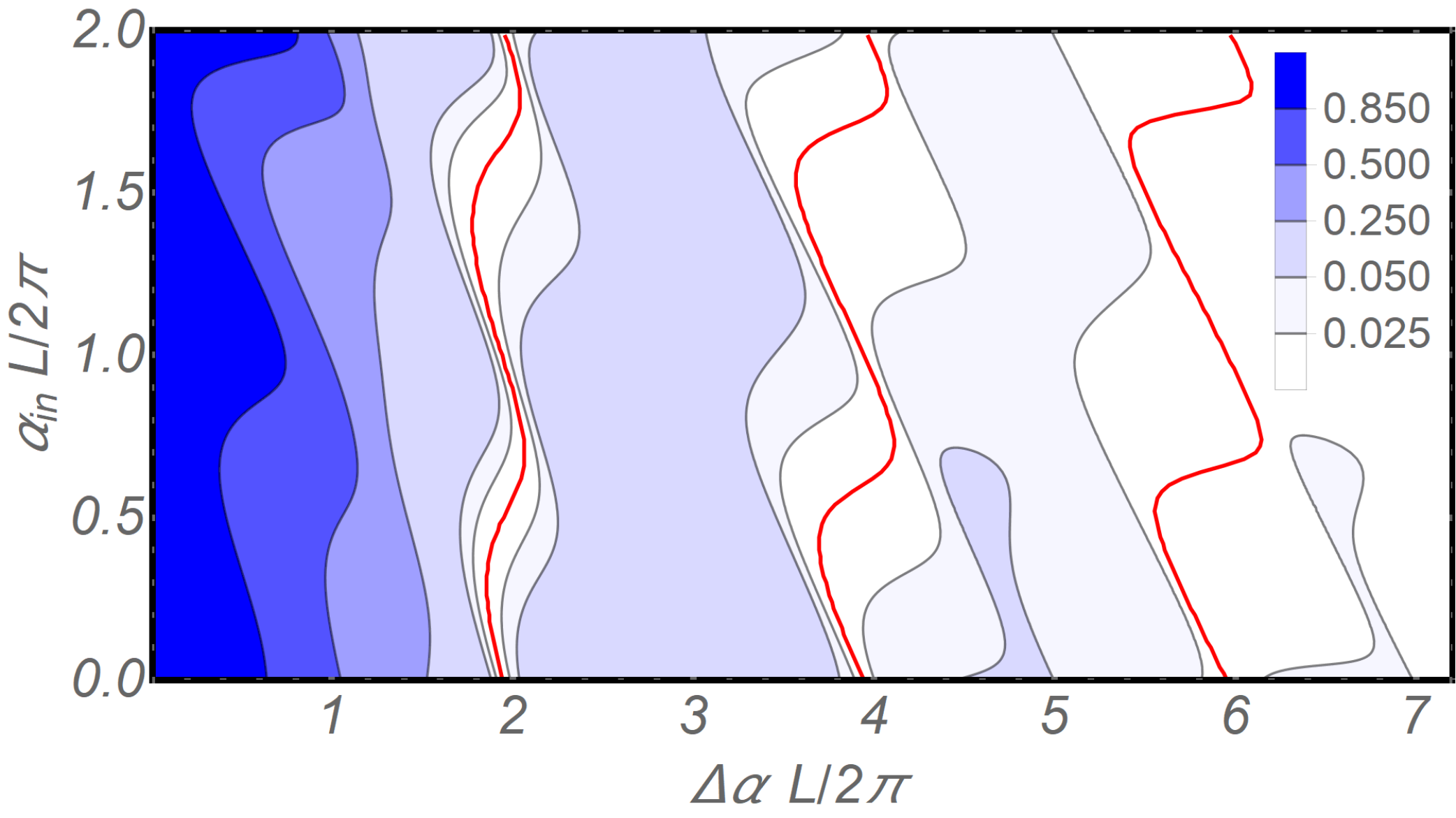}
\caption{The contour map of the amplitude $\abs{G(\alpha)}$ in unit of $\pi$ as a function of $\alpha_{in} L/2\pi$ and $\Delta\alpha L/2\pi$ where $\Delta \alpha =\alpha-\alpha_{in}$ and $L$ is the length of the nanoring. The orthogonality condition $\abs{G}=0$ is obtained at the zero Fisher $\alpha_n$ (red lines), at which a $\pi$-jump in the phase $\gamma$ occurs. The initial state is the ground state for $\alpha_{in}$, and the wire is an ellipse with the ratio of the semiaxis length set to $a/b=0.4$, so that level crossings with other energies do not occur in the interval under consideration. }
\label{fig:DQPT}
\end{figure}

This behavior can be related to the occurrence of a DQPT when the Rashba SO coupling is changed through a value $\alpha_n$.

%Indeed, the phase $\gamma$ is the argument of the Loschmidt amplitude $G(\alpha)=\braket{E_+(\alpha_{in})}{\Psi_+(\alpha)}$.

For a generic non-adiabatic evolution the Loschmidt amplitude $G(\alpha)=\braket{E_+(\alpha_{in})}{\Psi_+(\alpha)}$ describes how the evolved state deviates from the initial one, and can be employed to detect the occurrence of a DQPT when $G$ goes to zero at the so-called Fisher zeros for a given time instant associated to the spin-orbit coupling amplitude $\alpha_n$.
% at the so-called Fisher zeros for a given time interval associated to the spin-orbit coupling amplitude $\alpha_n$.
%When a value $\alpha_n$ is crossed, the amplitude $G$ goes through zero and the phase $\gamma$ jumps by $\pi$ (see Fig.~\ref{fig:DQPT}).
%Such behavior allows us to identify  DQPT at the Fisher zero $\alpha_n$.
%Then we can identify the value $\alpha_n$ as a Fisher zero.
In particular, for the analysis reported in Fig.~\ref{fig:DQPT} the critical values $\{\alpha_n\}$, indicated with red lines, are distributed in a non periodic way. We note that the amplitude $G(\alpha)$ crosses zero linearly when $\alpha$ is nearby the value $\alpha_n$.

\begin{figure}[!t]
%\centering
%\onefigure[width=80mm]{fig_3.pdf} %
\includegraphics[width=0.98\columnwidth]{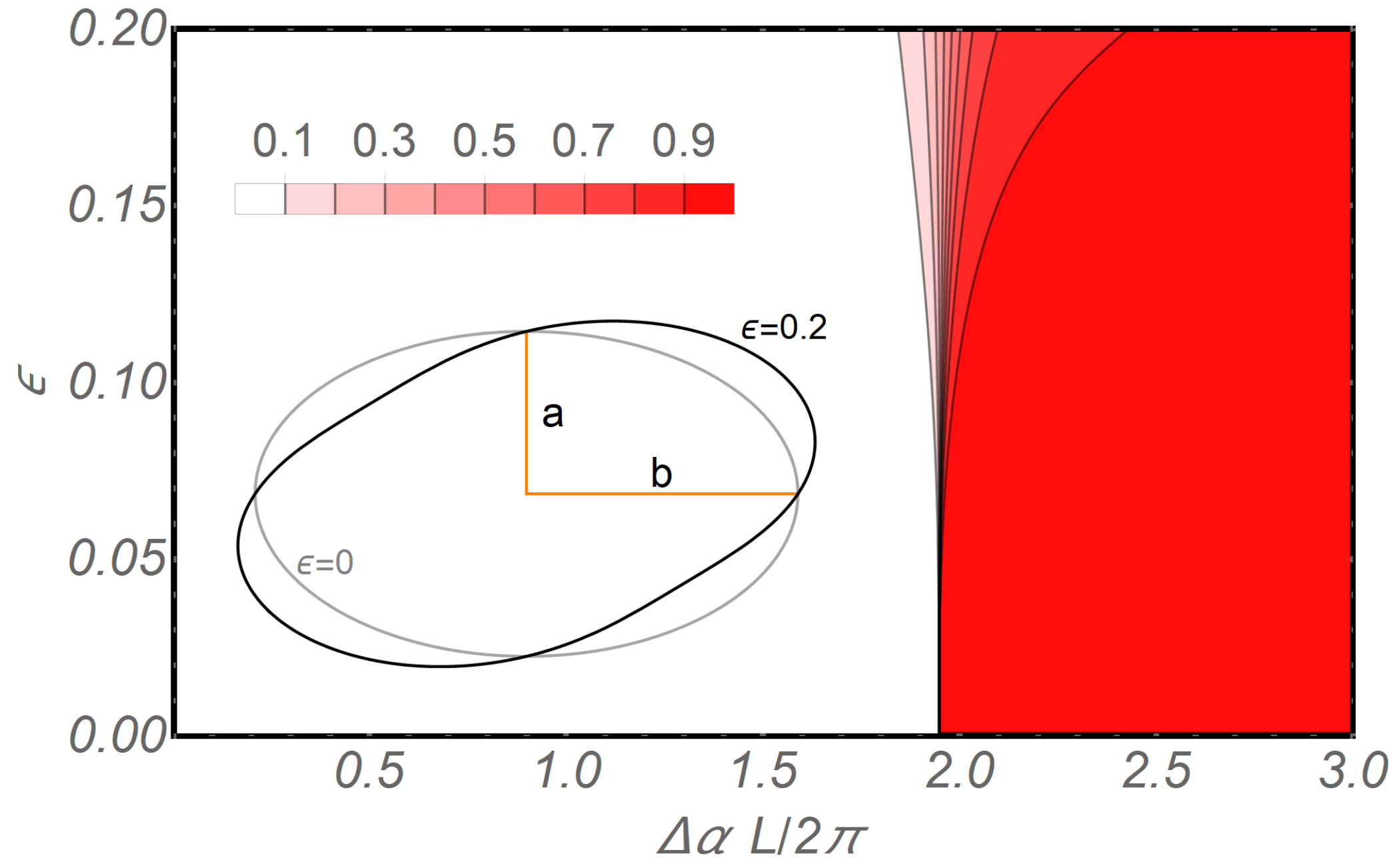}
\caption{Color map of the geometric phase $\gamma$ in unit of $\pi$ as a function of $\Delta\alpha L/2\pi$. A geometric non-Abelian phase is generated in the class $C_1\setminus C_2$ and a dynamical phase transition does not occur. We have considered a nanoring parameterized by $x=b (1+\epsilon \sin(4 t))\cos t$, $y=a (1+\epsilon \sin(4 t))\sin t$ with $t\in[0,2\pi]$. For $\epsilon=0$ the nanoring is elliptical. The results refer to $a/b=0.4$, and the initial configuration is the ground state at $\alpha_{in}=0$. }
\label{fig:na}
\end{figure}

Starting from the previous analysis, we observe that the reflection symmetry makes the phase $\gamma$ quantized so that it prevents the generation of a linear combination of the Kramers doublet. However, it is enough to consider a small modification of the spatial profile that breaks the mirror symmetry by keeping the inversion in order to get a geometric non-Abelian phase factor $U_g$ with $\gamma \neq n\pi$ and $n$ an integer. To demonstrate such effect, we apply a slight deformation to the elliptical shape of the nanoring by introducing a suitable parameterization of the coordinates (see inset in Fig.~\ref{fig:na}). As one can observe, by inspection of  Fig.~\ref{fig:na}, the change of the phase $\gamma$ in proximity of the critical amplitudes $\alpha_n$ becomes smooth when the nanoring is deformed away from the elliptical shape ({\it i.e.} $\epsilon=0$).

{\it Discussion and conclusions --} We show that a non-Abelian quantum manipulation and dynamical quantum phase transitions can be obtained by combining a variation of the Rashba SO coupling and the geometrical curvature in narrow semiconducting nanorings.

%The geometric and adiabatic manipulation of time-reversal symmetric %Kramers doublet in small nanorings can have a prospective impact for %designing innovative platforms for quantum manipulation especially for %the design of fault tolerant quantum gates~\cite{golovach10}.
Apart from the spatial confinement, a non-uniform curvature can gap the ground-state energy from the other levels, so that an adiabatic evolution in the ground-state sector can be achieved through a slow variation of the Rashba SO coupling.
In particular the gap can be generally expressed as  $\Delta E = g h^2/2mL^2$, where the factor $g<1$ depends on the geometric details of the nanostructure. We expect that the  adiabatic approximation for the time evolution of the ground state is suitable for $\eta = \hbar\abs{\bra{E_+'}\frac{\text{d} H}{\text{d} t} \ket{E_+}}/\Delta E^2\ll 1 $, where $E'$ is the first energy level and $\Delta E = E'-E$ is the energy separation.
Taking into account the expression of the gap $\Delta E$ then
$\eta = \frac{L^2}{\pi g \lambda_c c }\frac{\abs{\frac{\text{d}\alpha}{\text{d} t}}L}{2\pi} \abs{\bra{E_+'}\sigma_N \ket{E_+}}
$ with $\lambda_c =h/mc$ being the so-called Compton wavelength.
For a length $L\sim 60$ nm, an effective electron mass $m \sim 0.07 m_e$ and a geometric factor $g\gtrsim 0.1$, the adiabatic evolution in a range $\Delta \alpha L \approx \pi$ can be realized with a linear drive of duration $\Delta t \gg 10^{-12} \text{s}$.
The typical spin decoherence times in semiconducting nanostructures~\cite{hanson07} should then allow to observe such quantum evolution in laboratory. %exploit the holonomy for computation purposes.
Another relevant outcome of our study is that symmetric nanoring can realize an innovative quantum platform for achieving and characterizing a DQPT.
For the system upon examination, the DQPT can be experimentally detected by designing a nanoring tunnel coupled to a quantum dot.
We consider the case in which the Kramers states $\ket{E_\pm}$ are selected through an external perturbation $V$. The final state $\ket{\Psi_+(\alpha)}$ is the time evolved configuration of the initial ground-state $\ket{E_+(\alpha_{in})}$ after changing the Rashba SO coupling from $\alpha_{in}$ to $\alpha$. Then, at the end of the process we  perform a sudden change $\alpha\rightarrow \alpha_{in}$, so that the final state remains $\ket{\Psi_+(\alpha)}$. Hence, through the coupling with the quantum dot one can probe only the electron tunnelling in a given energy window around $E(\alpha_{in})$, {\it e.g.} the spectral density function is zero above the energy $E(\alpha_{in})$. Within the first order in the perturbation $V$ only transitions from $\ket{\Psi_+(\alpha)}$ into an eigenstate $\ket{E'_+(\alpha_{in})}$ are allowed if the element matrix $\bra{E'_+(\alpha_{in})}V \ket{\Psi_+(\alpha)}= V_{E'} \braket{E'_+(\alpha_{in})}{\Psi_+(\alpha)}$ is nonzero. However, at the critical values $\alpha=\alpha_n$, the Loschmidt amplitude is vanishing, {\it i.e.} $G=0$, so that the transition to the ground state occurs only through high order processes, thus leading to a sensible decrease of the tunnel current between the nanoring and the dot.

\end{document}